# Biotic Control of Earth's Climate:
# An Ecohydrological Perspective on the Phanerozoic Temperature Record

Allen Hunt[1] and Didier Sornette[2]

[1]Department of Physics, Wright State University
3640 Colonel Glenn Highway, Dayton, OH 45435, USA
[2]Institute of Risk Analysis, Prediction and Management, Academy for Advanced Interdisciplinary Studies, Southern University of Science and Technology, Shenzhen, China

**Abstract**
We propose a unified framework linking silicate weathering feedbacks, ecohydrological optimality and vegetation-climate interactions to explain Earth's Global Average Temperature (GAT) evolution over the Phanerozoic. The framework integrates two complementary processes: (i) solute-transport-limited weathering, which governs long-term carbon sequestration, and (ii) ecohydrological optimality, which constrains vegetation productivity through the balance between water availability and carbon uptake. Together, these processes establish two emergent climatic thresholds, around 15 °C and 33 °C, that define a stable corridor for the Earth system. Below 15 °C, declining evapotranspiration and rising albedo accelerate cooling; above 33 °C, heat and water stress suppress vegetation and limit further warming. Analysis of ten independent studies (2019-2024) confirms the recurrent appearance of these thresholds in both modern ecosystems and Phanerozoic temperature reconstructions. Major biological innovations, such as the colonization of land by plants, the rise of forests, and the emergence of $C_4$ photosynthesis, have reorganized these feedbacks, repeatedly shifting the planet between quasi-stable climate states. The resulting picture is one in which the biosphere acts as an active regulator of climate, modulating and at times overriding geochemical feedbacks. This framework unifies climate, tectonics, and life into a single ecohydrological system, offering a coherent explanation for the bounded variability of Earth's temperature across 500 million years and insight into its future resilience.

**Significance Statement**
Understanding how Earth's climate has remained within habitable bounds for half a billion years is central to addressing today's climate crisis. Our work reveals that vegetation and the hydrological cycle together act as a planetary regulator, constraining global temperature between two emergent limits, around 15 °C and 33 °C, that reflect the optimal range for ecosystem productivity. These limits arise from the balance between water availability, carbon uptake, and energy exchange at the land surface, linking biological processes directly to global climate stability. By integrating ecohydrology with silicate weathering feedbacks, we show that major evolutionary innovations, such as the rise of forests and $C_4$ plants, reshaped these feedbacks and drove shifts between quasi-stable climate states. This perspective challenges the view of life as a passive climate responder, positioning the biosphere as an active regulator. Understanding these self-stabilizing mechanisms offers critical insight into how Earth might respond, or fail to respond, to ongoing human-driven climate perturbations.

## 1. Introduction

A central unresolved problem in Earth system science is understanding why the planet's surface temperature has remained within habitable limits over the past 500 million years, despite major fluctuations in atmospheric $CO_2$, solar luminosity, and tectonic configurations. The Phanerozoic Global Average Temperature (GAT) record shows swings from roughly 8 °C to 33 °C, punctuated by both extreme warmth and profound glaciations. Classical carbon-cycle theory attributes these variations to changes in atmospheric $CO_2$, governed by the balance between volcanic outgassing and silicate weathering. In this view, first articulated by Urey (1952) and developed by Berner (1992), silicate weathering acts as a self-regulating "thermostat" that cools Earth when $CO_2$ concentrations are high and slows during cooler periods.

Yet, the Phanerozoic record reveals systematic departures from this simple thermostat behavior. The Late Ordovician (Hirnantian) glaciation (~450-440 Ma) and the Pleistocene ice ages both exceed the cooling expected from silicate weathering feedbacks alone. Mills et al. (2019) and Scotese et al. (2021) note that these events cannot be explained without invoking direct biological influences, echoing the earlier proposal by Lenton et al. (2012) that the Ordovician glaciation was triggered by the first land plants. Likewise, periods of exceptional warmth such as the Permian-Triassic interval cannot be reconciled with carbon-cycle models based solely on abiotic feedbacks. These discrepancies suggest that the long-term $CO_2$-weathering balance, while fundamental, does not by itself govern the amplitude or timing of major climate transitions.

A growing body of work indicates that water availability and transport are the dominant regulators of global weathering rates. Silicate weathering depends not only on temperature but critically on the flux of liquid water through soils and regolith, which controls both chemical reaction rates and solute export (Gibbs et al. 1999; Maher and Chamberlain 2014; Mills et al. 2019). In this sense, hydrologic flux and not temperature is the true driver of the long-term $CO_2$ sink. Because the biosphere exerts direct control on evapotranspiration and runoff through vegetation cover, rooting depth, and stomatal behavior, it follows that biological processes are inseparable from the regulation of global climate.

The apparent breakdowns of the $CO_2$-weathering feedback at key intervals such as the Hirnantian, the Cenozoic, and the Permian-Triassic, thus reflect moments when the coupling between life, water, and rock changed fundamentally. These shifts likely arose from evolutionary innovations that restructured the terrestrial water cycle, such as the colonization of land by plants, the rise of forests, and the emergence of water-efficient $C_4$ vegetation.

The goal of this study is to develop and test a unified ecohydrological framework that integrates silicate weathering feedbacks with vegetation-climate interactions. By coupling solute-transport–limited weathering to the principle of ecohydrological optimality, we show that this approach provides a parsimonious explanation for the major oscillations in Phanerozoic GAT. This framework captures how the coevolution of the hydrological cycle and the biosphere has bounded Earth's temperature within a narrow range of stability for much of its history and offers new insight into the mechanisms that continue to stabilize, or destabilize, the climate today.

## 2. Theoretical Basis

We re-interpret the Phanerozoic temperature record in terms of a comprehensive ecohydrological framework based on Earth's water balance. This approach builds on scaling theories for vegetation growth (describing how plant growth depends on time and evapotranspiration flux) and for chemical weathering and soil formation (describing their dependence on time and runoff-driven solute transport). By coupling these theories, we link the carbon cycle to the water cycle in a way that accounts for biological optimization and hydrologic constraints (Hunt et al. 2021; Hunt et al. 2020; Hunt and Sahimi 2017; Hunt and Manzoni, 2015).

## 2.1 Water Balance Framework and the ecohydrological optimality principle

Vegetation growth is fundamentally constrained by water availability. Net primary productivity (NPP) of plants increases with evapotranspiration (ET) – the flux of water transpired by plants and evaporated from land surfaces – which represents the proportion of precipitation effectively converted into biomass. In contrast, rock weathering and soil formation depends on the runoff component (Q), the fraction of precipitation that infiltrates and transports dissolved weathering products, thereby enhancing $CO_2$ sequestration in minerals. The partitioning of precipitation (P) into evapotranspiration (ET) and Q thus defines the water balance that underpins the global carbon cycle, aptly described as its "template" (NRC, 1991).

Our central postulate—the ecohydrological optimality principle—holds that terrestrial ecosystems have evolved to maximize net primary productivity (NPP) under the joint constraints of water and energy availability (Odum, 1959; Schymanski et al., 2006). In this view, vegetation self-organizes to optimally allocate precipitation between transpiration and runoff, thereby achieving the most efficient conversion of available water into biomass while maintaining the hydrological balance of the system. This ecohydrological optimality reconciles "bottom-up" physical constraints with "top-down" biological selection (Sivapalan, 2005; Schymanski et al., 2006), providing a unified framework that links hydrologic fluxes (*P*, *ET*, *Q*) to vegetation function and climate regulation. The bottom-up component specifies NPP as a function of hydrologic fluxes (precipitation (*P*), evapotranspiration (*ET*), and runoff (*Q*)), whereas the top-down component posits that plant communities modulate water use to maximize collective growth and reproductive success under prevailing conditions.

The ecohydrological optimality principle extends a lineage of foundational ideas in ecology and thermodynamics: Odum's (1959) concept of ecosystem maturity, Lotka's (1922) maximum power principle, and the optimal stomatal regulation framework of Cowan and Farquhar (1977). Each posits that biological systems self-organize to maximize useful energy capture and conversion within environmental limits. In the context of the terrestrial water cycle, this translates into an optimal partitioning of precipitation *P* into evapotranspiration *ET* and runoff *Q*, such that biomass production per unit of available water is maximized while maintaining hydrological balance.

This optimal allocation has a clear physical and physiological basis. The bottom-up constraints arise from radiative energy balance, soil-plant hydraulics, and atmospheric vapor-pressure deficit, which jointly determine potential *ET* and runoff. The top-down adaptive component reflects how vegetation structure, rooting depth, and stomatal conductance co-evolve to exploit available water efficiently while minimizing hydraulic risk. Optimal stomatal behavior—minimizing water loss per carbon gained—has been verified from the leaf to ecosystem scales (Medlyn et al., 2011), and globally, satellite and eddy-covariance data confirm that NPP rises nearly linearly with ET up to biome-specific thresholds (Zhang et al., 2016), understandable in terms of a fixed adaptation to smaller, within biome, climate variations. Across biomes, however, changing plant adaptation characteristics (e.g., root and/or canopy structure) is consistent with the empirical evidence of a stronger dependence on ET (Rosezweig, 1968; Budyko, 1973). These regularities indicate that vegetation operates near the theoretical efficiency limit set by ecohydrological constraints.

Over geological timescales, plant evolution has reinforced this principle. The colonization of land by early plants, the rise of deep-rooted forests, and later the emergence of water-efficient $C_4$ vegetation successively increased water-use efficiency and expanded the climatic envelope of productivity. Each innovation reorganized the coupling between carbon and water fluxes, suggesting that biotic evolution repeatedly re-optimized the global water balance to sustain productivity under changing climates. This long-term coevolution provides a macro-scale validation of the optimality postulate.

Importantly, the optimized water-balance framework has also shown predictive success across spatial scales, offering independent empirical corroboration.
1. **Net primary productivity across climate gradients.** *Hunt et al.* (2024) demonstrated that applying this optimality framework reproduces observed NPP across Asia's diverse climate zones with exceptional accuracy ($R^2 \approx 0.96$), consistent with Budyko's (1973) empirical climate–productivity relationship.
2. **Tree species richness and energy-diversity gradients.** *Hunt* (2025) extended the theory to explain continental-scale patterns of tree species richness across North America and Asia, capturing the canonical energy-diversity gradients observed by Currie (1991) and Latham & Ricklefs (1993).
3. **Streamflow sensitivity and hydrologic scaling laws.** The framework predicts how streamflow elasticity—the responsiveness of runoff to changes in precipitation—varies systematically with climate and drainage-basin area (*Hunt et al.*, 2023; 2025a), matching observed scaling behaviors in catchment hydrology. Predictions of continental-scale evapotranspiration likewise agree closely with empirical datasets (*Hunt et al.*, 2025b), further confirming the robustness of the underlying optimization principle.

Together, these theoretical arguments and quantitative tests provide compelling support for the notion that terrestrial ecosystems behave as self-optimizing systems. Through evolutionary adaptation and physiological regulation, they adjust their water use to maximize productivity and stability within environmental limits. The success of this framework in reproducing patterns of NPP, biodiversity, and hydrologic response across temporal and spatial scales establishes ecohydrological optimality as a unifying principle linking biological evolution, climatic forcing, and hydrological dynamics, from local ecosystems to the Phanerozoic-scale regulation of Earth's carbon-water cycle.

**2.2 Transport-based Chemical Weathering and Soil Formation**
The theoretical treatment of chemical weathering used here has not yet been incorporated into existing climate models, even though its key outcomes have sometimes been included in models in a piecemeal, ad hoc fashion. Clarifying this theory is important for two reasons: **(1)** to better understand the derivation of the water balance optimality described above, and **(2)** to show how a more realistic depiction of chemical weathering can improve our qualitative and quantitative understanding of past climate-change mysteries.

Central to this theory is the role of runoff in governing weathering rates. A series of publications (Yu and Hunt 2017a, 2017b, 2017c) demonstrated that field-measured chemical weathering rates are essentially controlled by the rate at which water (carrying dissolved products) can transport solutes away from the weathering front. In other words, the slow transport of solutes through soils and rocks is the limiting factor, not the intrinsic reaction kinetics under laboratory conditions. In particular, this transport-limited theory predicts that weathering rates decay as a power law over time – dropping by about six orders of magnitude from timescales of weeks to timescales of millions of years – exactly as observed by (White and Brantley, 2003) in their compilation of weathering data. The model also predicts that weathering rates are proportional to runoff (water flux), consistent with earlier suggestions by (Gibbs et al., 1999) that global carbonic weathering rates scale with river discharge. These results hold over ten orders of magnitude of flow speeds (Maher 2010; Salehikhoo et al. 2012) and, under steady-state conditions, the weathering rate is further predicted to be proportional to the erosion (denudation) rate of the landscape – a link that was noted empirically by (Mills et al., 2019) as well.

The near-ubiquity of non-Gaussian solute transport in natural porous media (Cushman and O'Malley, 2015) underpins this behavior. Disordered soils and rocks do not conduct water evenly like a simple pipe; instead, flow follows a network of paths whose conductances vary widely. Tracer studies show characteristics of anomalous (non-Fickian) dispersion, such as a power law relationship between travel time and distance (with exponents ~1.5–2.0) (Berkowitz and Scher, 1995; Schlesinger et al. 1991) and a tendency for classical (Gaussian) models to underestimate both short- and long-time solute fluxes (Cortis & Berkowitz, 2004). In practical terms, this means that dissolved weathering products take a long time to exit the system, and this transport lag causes weathering reactions to slow down over time even if fresh mineral surface is continually generated. Models that neglect such non-Gaussian transport limitations often invoke other reasons for the slowing of weathering (like depletion of easily weatherable minerals or formation of protective coatings), but in our framework the slowdown emerges inevitably from transport limitations.

Overall, our transport-based weathering theory offers a parsimonious and accurate explanation for several key observations: the time-dependent decay of weathering rates (in both lab and field, e.g., White and Brantley, 2003), the proportionality of weathering rate to water flow rate (Maher, 2010; Salehikhoo et al. 2012), and the coupling of weathering and denudation rates (DiBiase et al. 2012; Mills et al. 2019). This simplicity and breadth of applicability make it a strong candidate for inclusion in Earth system models. We now summarize the theoretical formulation and its consequences.

Porous media such as soil, regolith, or fractured rock can be viewed as disordered networks (Hunt & Sahimi 2017, 2024). Water flow in such networks is controlled by the lowest-conductance "bottleneck" pathways, as captured by critical path theory, a percolation-inspired framework introduced by Ambegaokar, Halperin, and Langer (1971) and Pollak (1972) and later developed by Sheng (1980, 1983) and Hunt (2001). In this view, the effective hydraulic conductivity is determined by the critical conductance at which a spanning (infinite) cluster first forms: the so-called critical path, corresponding to the percolation threshold of the network. The time required for a solute to travel a given Euclidean distance through the porous medium does not scale linearly with distance $x$, but instead follows a power law relationship governed by the fractal geometry of the percolating flow paths. In a disordered porous network, solute transport occurs along the backbone of the percolation cluster, which is the subset of connected pores that sustain flow. The process is advective rather than diffusive: solute particles move under a nearly constant velocity field along these connected channels, so that travel time is governed by the geometric elongation of the flow path rather than by molecular diffusive dispersion. While the local velocity along the backbone remains approximately constant, the total path length increases faster than the Euclidean distance x, reflecting the tortuous and self-similar nature of the percolating flow network. Denoting the fractal dimension of this backbone as $D_b$ >1, the total time to reach a distance x scales as $x^{D_b}$, and hence the most probable travel time satisfies $t \propto x^{D_b}$. The discoverers of this relationship (Lee et al., 1999) refer to it as a "temporal tortuosity." In three-dimensional flow through porous media, $D_b \approx 1.87$ (Lee et al., 1999) – a reflection of the twisted, tortuous pathways water takes – whereas in effectively two-dimensional flow (e.g., along a planar fracture) $D_b \approx 1.64$.

To turn this proportionality into an equation, we introduce a characteristic length $x_0$ (on the order of a median grain or pore size) and a characteristic time $t_0$ (associated with flow of velocity $v_0 = x_0/t_0$ across that length). Although the solute velocity $v_s$ must decline with increasing length scale, at the scale of a representative pore, it equals the mean percolating water velocity $v_0 = (P - ET)/\varphi$ (where P is precipitation minus evapotranspiration ET, divided by porosity $\varphi$). The term P-ET represents the net infiltration flux, which is the portion of precipitation that percolates through the ground, while division by φ converts this bulk (Darcy) flux into the mean velocity within the connected pore space. The actual water velocity within the pores is higher than the bulk flux P-ET because only a fraction φ of the medium's total cross-sectional area is available for flow. This represents the portion of precipitation that percolates through the ground.

The resulting relationship for transport distance $x$ as a function of time $t$ is:

$$x = d_{50} \left[\frac{t}{t_0}\right]^{\frac{1}{D_b}} = d_{50} \left[\frac{(P-ET)}{\varphi\, d_{50}} t\right]^{\frac{1}{D_b}} \quad \text{with } t_0 := \frac{d_{50}}{v_0} \text{ and } v_0 = \frac{P-ET}{\varphi} \quad (1)$$

where $d_{50}$ is the median grain diameter (or median pore spacing), serving as the characteristic microscopic length scale of the medium and $D_b \approx 1.87$. This expression (1) encapsulates the sublinear growth of transport distance $x$ with time $t$ which is a result of the fact that flow paths become increasingly tortuous at larger scales. Moreover, the dependence of transport distance $x$ on water flow velocity $(P - ET)/\varphi$ is also to the power $1/D_b \approx 1/1.87 = 0.53$, close to ½.

Differentiating (1) with respect to time yields the solute velocity $v_s(t)$, representing the speed at which weathering products are transported away from the reaction front at time $t$:

$$v_s(t) = \frac{1}{D_b} \frac{d_{50}}{t} \left[\frac{t}{t_0}\right]^{\frac{1}{D_b}} \quad (2)$$

Efficient removal of these products is essential; if $v_s(t)$ is too low, local equilibrium is reached, and the weathering reaction effectively ceases. Equation (2) yields $v_s(t) \propto t^{1/1.87 - 1} = t^{-0.47}$. Let us consider a characteristic soil or regolith thickness $z$ as the transport distance that corresponds to a typical transport time $t_z = t_0 \left(\frac{z}{d_{50}}\right)^{D_b}$. Substituting this time $t_z$ into $v_s(t)$ (2) yields

$$v_s := v_s(t_z) = \frac{1}{D_b} v_0 \left(\frac{d_{50}}{z}\right)^{D_b - 1} = \frac{1}{1.87} v_0 \left(\frac{d_{50}}{z}\right)^{0.87} \quad (3)$$

where $v_0$ defined in (1) is the mean percolating water velocity, and $D_b = 1.87$ for 3D flow.

Multiplying $v_s$ by the molar density $C_0$ of weathering products (number of moles of dissolved weathering products per unit volume of pore water), we obtain the areal flux $J$ of weathering products exported from the weathering zone thickness $z$, This expression quantifies the rate at which dissolved products are removed through a soil column of thickness $z$. It shows that the flux decreases sublinearly with increasing $z$, because transport becomes slower as pathways grow more tortuous.

Finally, inverting this relation (3) to express $z$ as a function of $v_s$ (or equivalently $J$) gives

$$z = d_{50} \left(\frac{1}{D_b} \frac{v_0}{v_s}\right)^{1/(D_b-1)} = d_{50} \left(\frac{1}{D_b} \frac{P-ET}{\varphi v_s}\right)^{1/(D_b-1)} \quad (4)$$

This equation has a key implication: it links the steady-state soil thickness to the balance between solute export velocity $v_s$ (or flux $J$) and the mean percolating water velocity $v_0$.

Now, because the solute transport distance $x(t)$ (1) corresponds, at steady state, to the soil thickness $z$ maintained by a balance between weathering and erosion, the same process that exports solutes also governs soil production. The rate of soil production, denoted $P_s$, must therefore be proportional to the velocity with which the weathering front advances downward, that is, to the solute export velocity $v_s(t_z)$:

$$P_s \propto v_s(t_z) = \frac{1}{D_b} v_0 \left(\frac{d_{50}}{z}\right)^{D_b - 1}. \quad (5)$$

Equation (5) thus defines the soil production function: it decreases with increasing soil thickness because transport slows as pathways lengthen and become more tortuous. This expression provides the physical link between chemical weathering kinetics, solute transport, and the steady-state soil depth.

Since the solute transport distance $x(t)$ in Eq. (1) can be interpreted, at steady state, as the soil or regolith thickness $z$, its time derivative $v_s(t) = \mathrm{d}x/\mathrm{d}t$ from Eq. (2) is not only a measure of solute transport but also directly linked to the rate at which fresh material is transformed into soil. We have therefore defined the soil production function as being proportional to the solute export velocity $v_s(t_z)$ from Eq. (3). Once weathering products are removed from the parent material, the remaining solid mass is reduced; however, the large increase in porosity during rock-to-soil transformation lowers the bulk density, so that the volume of soil produced typically exceeds the volume of the original bedrock. Consequently, converting the solute transport velocity $v_s$ into an equivalent rate of surface lowering (soil production per unit area) requires a density correction factor that depends on the densities of both the parent rock and the resulting soil.

Under steady-state conditions, soil production balances soil loss through erosion or denudation, characterized by a rate $D$. In that case, the soil thickness remains constant, and $D$ becomes proportional to the solute transport velocity $v_s$:
$$D \propto v_s. \qquad (6)$$
It is important to emphasize that these two quantities are related but not identical: $v_s$ is the internal velocity of solute migration within the pore network, whereas $D$ represents the macroscopic rate of soil removal from the surface. Their proportionality reflects the fact that efficient solute export sustains ongoing weathering and, hence, continuous soil generation.

Replacing $v_s$ by the denudation rate $D$ in Eq. (4) yields the following steady-state relationship between soil thickness $z$, water flux $(P - ET)/\phi$, and denudation rate $D$:
$$z \propto d_{50} \left(\frac{1}{D_b}\frac{v_0}{D}\right)^{1/(D_b-1)} = d_{50}\left(\frac{1}{D_b}\frac{P-ET}{\phi D}\right)^{1/(D_b-1)}. \qquad (7)$$
This expression predicts that thicker soils develop where infiltration fluxes $(P - ET)$ are higher or denudation rates $D$ are lower.

These theoretical predictions have been repeatedly validated. They reproduce the observed evolution of soil profiles over time and depth (including known denudation rates) (Yu & Hunt 2017a,b,c; Egli et al. 2018). They are consistent with field compilations of weathering rates (White and Brantley, 2003), and align with field and laboratory data showing weathering rates proportional to flow rates (Maher, 2010; Salehikhoo et al. 2012). Remarkably, all these successful comparisons derive from the same underlying transport law (3), demonstrating excellent predictive capability without any parameter tuning.

In practical terms, soil thickness exhibits only a weak dependence on climate. This arises because variations in the water flux term (P – ET, i.e., runoff) are largely offset by corresponding adjustments in the weathering/denudation rate $v_s$, leading to an overall buffering of soil depth against climatic fluctuations. Specifically, both P – ET and $v_s$ tend to increase with precipitation P (Reiners et al. 2003), largely canceling each other out in expression (4). Except in very humid regions, ET is a consistent fraction of P, and in general denudation rates are roughly proportional to P – ET, as well. Instead, soil thickness is much more strongly influenced by topography: Denudation rates D rise sharply with terrain relief, diverging as slopes approach a critical angle (Montgomery & Brandon, 2002). Thus, as slopes steepen, soils become thinner, eventually leading to bare rock or landslide-prone hillslopes as predicted by the model as shown by expression (7) (Yu et al., 2019). In short, climate factors like precipitation have a modest effect on long-term soil development, whereas tectonic factors (slope/relief) are dominant in controlling soil depth. Moreover, the proportionality of $v_s$ to both $v_0$ and to $D$, tends to restrict the variability in concentrations of weathering products in runoff.

Another way to determine when intrinsic reaction kinetics (rather than transport) control weathering is through a Damköhler number analysis. The Damköhler number ($Da_I$) is a dimensionless ratio that compares the characteristic timescales of transport and chemical reaction. It quantifies which process, solute transport or mineral dissolution, acts as the rate-limiting step. Formally,

$$Da_I = \frac{\tau_{ad}}{\tau_r}, \qquad (8)$$

where $\tau_{ad}$ is the advective transport time (the time required for solutes to traverse the soil or regolith column), and $\tau_r$ is the intrinsic reaction time (the time required for dissolution reactions to proceed at the mineral surface under laboratory conditions).

- If $Da_I > 1$, the transport time is longer than the reaction time, implying that solute advection and diffusion limit the overall weathering rate.
- Conversely, if $Da_I < 1$, chemical reactions are slower than transport, and reaction kinetics dominate.

Following Yu and Hunt (2017c) and Salehikhoo et al. (2012), the advective time scale $\tau_{ad}$ is obtained from the transport law (1) by replacing the transport distance with the soil depth $z$:

$$\tau_{ad} = \frac{d_{50}}{v_0} \left(\frac{z}{d_{50}}\right)^{D_b}, \qquad (9)$$

where $v_0 = (P - ET)/\phi$ is the mean pore velocity and $D_b \approx 1.87$ characterizes the fractal tortuosity of flow paths as before.

The intrinsic reaction time $\tau_r$ is related to the intrinsic dissolution rate $R$ (moles $m^{-2} s^{-1}$) by

$$\tau_r = \frac{V_p C}{RA}, \qquad (10)$$

where $V_p$ is the volume of a representative pore, $C$ is the molar concentration of reactive minerals (moles $m^{-3}$), and $A/V_p$ is their surface-area-to-volume ratio. This expression arises because $RA$ gives the rate of reaction per pore volume (mol $m^{-3} s^{-1}$), so the inverse defines the corresponding reaction time.

Combining these definitions yields the Damköhler number:

$$Da_I = \frac{\tau_{ad}}{\tau_r} = \frac{\left(\frac{d_{50}}{v_0}\right)\left(\frac{z}{d_{50}}\right)^{D_b}}{\frac{V_p C}{RA}} = \frac{\left(\frac{d_{50}}{v_0}\right)\left(\left[\frac{(P-ET)}{1.87 \phi D}\right]^{\frac{1}{D_b-1}}\right)^{D_b}}{\frac{V_p C}{RA}} \qquad (11)$$

Using these representative parameters (soil depth $\approx 1$ m, $v_0 \approx 0.5$ m yr$^{-1}$, $d_{50} \approx 30$ μm, $R \approx 10^{-13}$ mol m$^{-2}$ s$^{-1}$, $A/V_p \approx 10^5$ m$^2$ m$^{-3}$, $C \approx 10^4$ mol m$^{-3}$), Yu and Hunt (2017c) showed that within only a few years, corresponding to solute penetration of about 1 cm into unweathered rock, the Damköhler number exceeds unity ($Da_I > 1$). Beyond this shallow zone, solute transport rather than reaction kinetics controls the overall weathering rate for most silicate systems. Only in cases of extremely high relief and erosion (where soils are just a few centimeters or less thick) might weathering proceed fast enough that chemical kinetics (lab rates) remain relevant (Egli et al. 2018). The model captures this via an explicit factor $D^{-\beta}$ (with $\beta = D_b/(D_b-1) = (1.87)/(1.87-1) = 2.15$ for 3D flow) in the advection time. Because denudation rate $D$ itself increases dramatically as slopes approach a threshold (Montgomery & Brandon, 2002), there could be an abrupt crossover to kinetics-limited weathering at those extreme conditions. Generally, though, for most landscapes and timescales of interest, we expect weathering rates in nature to be limited by transport of solutes rather than by the intrinsic reaction speeds.

For clarity, we note that some previous treatments have misinterpreted the interplay of transport and kinetics. For example, Salehikhoo et al. (2012) computed a Damköhler number without accounting for the slowing of transport with time, and they inadvertently used "field" reaction rates (already reduced by transport limitations) in place of true lab rates. This effectively inverted the diagnosis, making it seem as though kinetics were limiting when in fact transport was the culprit. Our framework avoids that pitfall by properly including the time-dependence of transport from the outset.

### 2.3 Vegetation Growth and Productivity

Roots, the subterranean foraging organs of plants, have long been recognized to exhibit fractal-like architecture (Lynch, 1995). This is not merely a curious geometric fact as root architecture critically influences plant productivity because soil resources (water, nutrients) are patchily distributed. As Lynch (1995) observed, "the spatial deployment of the root system will in large measure determine the ability of a plant to exploit those resources."

An analogy can be drawn between root growth in soil and the optimal flow paths discussed above for water. We hypothesize that roots explore the soil by following "optimal paths" of least resistance, akin to the optimal pathways in a percolation network. In simple terms, plant roots do not need conscious strategy to do this – rather, roots grow in response to water potential gradients, and water naturally travels along the easiest (highest conductance) paths. Thus, roots will tend to extend along channels where water (and nutrients) are most readily available. When an extending root encounters a nutrient-rich zone, it branches and proliferates there, effectively tracing back along those optimal paths to exploit the resource. This behavior allows the plant to gain the most nutrients and water for the least energy investment, a clear competitive advantage (Hunt and Manzoni, 2015).

Mathematically, one can treat root expansion similarly to solute transport. The original scaling predictions for root lateral growth are analogous to those for soil formation, differing mainly in the exponent related to dimensionality. The characteristic velocity becomes the pore-scale water flow rate, and the characteristic time the time between pore-scale root growth events. Later refinements amount to upscale time to the scale of an annual growing season and water flux to the annual transpiration rate, without changing the form of the prediction (Hunt et al. 2017). The resulting model for the lateral spread of roots *rls* (which for trees corresponds to canopy size or height up to ~40–50 m, Hunt and Manzoni, 2015) is:

$$rls = T_g \left(\frac{t}{t_g}\right)^{\frac{1}{D_{opt}}} \qquad (12)$$

where $T_g$ is the annual transpiration volume (water use), $t_g$ is time (in years of growth), and $D_{opt}$ is the optimal path exponent from percolation theory. The optimal path is similar to the chemical or shortest path in the Percolation literature, but it corresponds rather to the overall smallest resistance rather than shortest length. In a predominantly two-dimensional exploration (e.g., shallow lateral spreading near the soil surface), $D_{opt} \approx 1.21$. Using long-term data for tree growth, (Hunt and Manzoni, 2015; Hunt et al. 2020; etc.) found that this model with $D_{opt} = 1.21$ (the 2D value) accurately describes dozens of tree species' growth trajectories. In a minority of cases, an exponent near 1.43 was observed, consistent with the 3D optimal path value (1.43 is the 3D analog of 1.21 in 2D).

Further evidence for this root-growth scaling comes from allometry. Hunt and Manzoni (2015) showed that the classical allometric scaling exponent relating tree height to trunk diameter (~2/3; Enquist et al. 2000) should be modified to $2/(3 D_{opt})$ when root optimality is considered. For $D_{op}$ = 1.21 (shallow, 2D-like rooting), the predicted exponent is ~0.551; for $D_{op}$ = 1.43 (deep, 3D rooting), it is ~0.466. Empirical data compiled by Feldpausch et al. (2011) on tropical forest trees show a global height–diameter exponent of ~0.55, very close to the 2D optimal prediction. At continental scales, most datasets cluster around either ~0.55 or ~0.47, with only a few cases near 0.67 (which would imply no strong root-limit effect). Notably, South American tropical forests show a lower exponent (~0.47) consistent with deeper 3D rooting on that continent, whereas African forests show values closer to 0.67 in some regions (perhaps indicating other limiting factors). Overall, the allometric evidence supports the relevance of both 2D and 3D root optimization models in nature, with shallow-rooted (tortuosity-constrained) growth being more common globally, and deep-rooted strategies emerging in certain environments.

In applying our water balance theory, we make two simplifying approximations (Hunt et al. 2021): we treat subsurface runoff as effectively equal to total runoff $Q$, and transpiration as approximated by total evapotranspiration $ET$. These assumptions are supported by previous studies at large scales (since most runoff in steady state eventually infiltrates at some point, and ET includes the bulk of transpired water), but they can introduce some uncertainty for specific catchments. We assume these approximations are acceptable for broad, global-scale analysis.

**2.4 Ecohydrological Water Balance**
Using the scaling results for weathering (section 2.2) and for root growth (section 2.3), we summarize a unified theory of the water balance that couples the carbon and water cycles. The idea is to determine how water partitioning would be adjusted by ecosystems to maximize biomass production as measured by net primary productivity (NPP), given the constraints of water input P split between soil production (run-off) and plant uptake proxied by evapotranspiration (ET).

Empirical precedent for this strategy comes from classic studies of NPP vs. ET. Rosenzweig (1968) found that, across terrestrial biomes, NPP increases with annual ET according to $NPP \propto ET^{1.69}$. Hunt (2016) expanded this analysis to 20 studies (including Rosenzweig's data) and found $NPP \propto ET^{1.87}$. The exponent describing biomass production in terms of ET, ~1.87 is close to the 2D percolation mass fractal dimension of 1.9. Consistent with the extracted power, in the compiled data, ET ranged from ~20 mm yr$^{-1}$ in the Namibian desert to ~1600 mm yr$^{-1}$ in tropical rainforests and savannahs – a factor of 80 – while NPP over those biomes spanned from ~0.5 to ~2000 g C m$^{-2}$ – a factor of ~4000, consistent with $80^{1.9}$ = 4129. This suggests that the productivity of ecosystems is largely governed by how effectively water is used (transpired), which is consistent with a 2D optimality scenario for root systems. There is some evidence of a 3D pattern in a subset of the data: Bruce Milne (pers. comm.) noted that a fraction of sites follow a steeper slope ~2.5, matching the 3D percolation dimension. But globally, 1.9 appeared to dominate (Hunt, 2016).

Given the above, we allow for two modes in our model: a "2D" shallow-root optimality and a "3D" deep-root optimality. In the 2D mode, root systems spread laterally with fractal dimension ~1.9, and in the 3D mode they spread more isotropically with fractal dimension ~2.5. Thus,
$$NPP \propto ET^{d_f} \qquad (13)$$
with the predicted $d_f$ value equal to the mass fractal dimensionality from percolation theory (Stauffer and Aharony, 2003), which is proposed to correspond to the 2D value of 1.9 for shallow-rooted plants and to the 3D value of 2.5 for deep-rooted plants. It turns out that assuming either mode, 2D or 3D (for both root growth and root mass distribution), is reasonably supported by data. For example, Levang-Brilz and Biondini (2002) measured scaling exponents of root mass vs. lateral spread for 55 species and found values clustering near 1.79 or 2.65 for grasses (two groups) and ~2.5 for forbs. These correspond reasonably to the theoretical 2D (≈1.9) and 3D (≈2.5) values.

We now maximize NPP with respect to the water balance. Using the results from sections 2.2 and 2.3, total plant biomass (or productivity) can be expressed as a function of evapotranspiration ET and runoff (or infiltration) Q, the latter being the fraction of precipitation that infiltrates and transports dissolved weathering products. For a fixed precipitation $P$, these two fluxes are related by $Q = P - ET$. NPP (and hence total root biomass) depends on both:
- $ET$, because it controls carbon uptake through photosynthesis; and
- $Q$, because it controls soil depth and nutrient supply, via the dependence of soil thickness $z$ on infiltration given by Eq. (7).

Following Hunt et al. (2021), the total root mass $M$ is obtained by integrating the root mass contained in horizontal soil slices from the surface down to the full soil depth $z$. Each slice contributes a two-dimensional root mass proportional to $ET^{1.9}$ (from Eq. 13), and the number of such slices scales with $z$, which in turn depends on $Q = P - ET$. Substituting $z(Q)$ from Eq. (7) gives the composite dependence:
$$M = k\, ET^{1.9}(P - ET)^{1.15} \qquad (14)$$
where $k$ is a proportionality constant that includes geometric and density factors and the exponents correspond to the 2D case. The exponent 1.15 arises from the fractal scaling (7) of soil depth with infiltration flux, $z \propto (P - ET)^{1/(D_b - 1)}$, where the backbone dimension of the percolation network is $D_b \approx 1.87$, giving $1/(D_b - 1) \approx 1.15$. This function represents the total below-ground biomass (and thus productivity) as a balance between water use for transpiration and water infiltration sustaining nutrient transport. Maximizing M (hence NPP) with respect to ET (d*M*/d*ET*=0) yields the optimal partitioning of precipitation into evapotranspiration and runoff that maximizes total biomass and NPP..

For the 2D case (14), the result of the optimization is: ET = (1.9/(1.9+1.15)) P = 0.623 P, i.e. ~62.3% of precipitation is transpired at optimum. Equivalently, runoff Q = 0.377 P in this optimum state. In more intuitive terms, the ecosystem achieves its maximum productivity when it uses a bit less than two-thirds of incoming moisture for growth, with the rest becoming runoff.

In 3D, total root biomass $M$ must have the physical dimension of a volume (length³). Empirically and in models, root architectures are fractal with an effective mass (fractal) dimension $D_r \approx 2.5$ (between a sheet, $D = 2$, and a space-filling volume, $D = 3$). The baseline proportionality (13) thus scales as a length to the power $D_r$. If we were to multiply that directly by $z$ (length¹) to account for depth, we would obtain $length^{D_r+1} = length^{3.5}$, which is dimensionally too large for a 3D mass. To enforce dimensional closure (mass ~ length³), the depth contribution must supply exactly the missing exponent $\beta = 3 - D_r \approx 3 - 2.5 = 0.5$, so the depth factor is $z^\beta = z^{0.5}$, not $z^1$. Intuitively, this reflects that a $D_r \approx 2.5$ root network is already largely space-filling laterally; increasing depth packs additional roots sublinearly, hence $z^{0.5}$ (Hunt et al., 2021). Performing the optimization then yields ET = 0.813 P for the 3D-optimal case.

In summary, our model predicts two characteristic optimal ET/P ratios for ecosystems: α = 0.623 for predominantly shallow-rooted (2D) systems, and α = 0.813 for deep-rooted (3D) systems such that α may be represented as a dimensionally-dependent constant, α(d) with α(2d) = 0.623 and α(3d) = 0.813.

We emphasize that the above results apply only under conditions of equal water and energy supply, with water supply P neither so low as to limit ET, nor so high as to exceed evaporative demand. We consider next these cases of climatic constraints on the water balance.

For very humid (energy-limited) systems where P ≫ PET (potential ET), plants cannot evaporate all the rainfall because of energy constraints. Hunt et al. (2021) assumed that, in such cases, the optimality applies to P only up to the PET level. Beyond that, extra precipitation simply becomes runoff. Thus, if P > PET, we take $ET_{opt} \approx \alpha(d)$ (PET), and Q takes the remainder. Conversely, for arid (water-limited) systems (PET/P > 1), vegetation tends not to fully cover the ground. In those regions, Hunt et al. (2021) postulated (with quantitative support from Yang et al. 2009) that the fraction F of land covered by plants is F = P/PET. Where vegetation exists, the plants are assumed to partition the water exactly as under humid conditions, while on the uncovered fraction all rain evaporates directly. Thus, ET = α(d) P F + P (1 − F). Both constraints are simplifications (particularly, our treatment of water-limited systems may have shortcomings as it would predict increasing run-off with increasing vegetation coverage), but they serve as reasonable first-order corrections to the optimality theory from climate constraints. They ensure, for instance, that our model does not predict >100% of PET being used or ET > P in any scenario. In summary,

$$ET = \alpha(d)PET \qquad \frac{PET}{P} < 1 \qquad (15a)$$
$$ET = P\left[1 - (1 - \alpha(d))\frac{P}{PET}\right] \qquad \frac{PET}{P} > 1 \qquad (15b)$$

Equations (15a) and (15b) (in the original notation) encapsulate these constraints.

Hunt et al. (2023) analyzed global data and found that real-world ecosystems tend to operate between these two theoretical limits. For instance, observations of forests and grasslands around the world scatter between the α = 0.623 curve and the α = 0.813 curve (with no region showing a systematic deviation beyond this envelope). In fact, on large scales, the shallow-root optimum (α ≈ 0.62) appears to dominate: the observed water balance for continents closely matches the α = 0.623 prediction (Hunt et al. 2025b). This correlates with our earlier note that the global allometric exponent (~0.55) and the ET dependence of NPP (Hunt, 2016) indicated a 2D root influence; it seems the Earth's land vegetation in aggregate behaves in line with the 2D optimal water-use strategy. Moreover, by combining ET = 0.623 P with the NPP ∝ $ET^{1.9}$ relationship, Hunt et al. (2024b) showed that a single universal photosynthetic efficiency factor can predict the absolute NPP values of different climate zones (the classic Budyko belts) with $R^2$ = 0.96. This result means that, if we assume ecosystems globally are tuned to use ~62% of precipitation for growth, we can accurately calculate how productive each climate zone is, using just climate data. Additional confirmation comes from derived quantities: using α = 0.623 in the water balance, the model correctly predicts how streamflow elasticity varies with P/PET, storage changes, and Q (Hunt et al., 2023) and with basin area (Hunt et al., 2025a).

It is worth noting that the 3D optimal solution yields a lower NPP than the 2D solution under the same total P. Specifically, at PET/P = 1, the NPP for the 2D case is about 25% higher than for the 3D case. The 2D strategy benefits from a greater contribution of soil formation (Q) to NPP, whereas the 3D strategy channels more water into ET but with diminishing returns due to deeper roots having to invest more energy for less incremental gain as well as a shallower soil for carbon storage. This implies that ecosystems would prefer the shallow-root (2D) solution unless there are other advantages conferred by deep roots. Indeed, there are at least two such advantages:
1. In seasonal climates with long dry periods (especially tropical savannas or seasonal forests), deeper roots allow plants to tap moisture stored in deeper soil during droughts, conferring resilience and access to water when the shallow soil is dry.
2. Deep roots also enhance moisture recycling – the transpired water that goes back into the atmosphere can fall again as rain. In tropical regions, a higher fraction of rainfall comes from recycled water when forests have deep rooting systems (Makarieva et al., 2007). This can expand the zone of rainfall inland (the "biotic pump" hypothesis) and is thought to be important for explaining past greenhouse climates in models (DeConto et al., 1999) through meridional atmospheric moisture and latent heat transport.

From these considerations, we expect that, in times or places where large land areas are warm and humid (e.g., extensive tropical supercontinents), vegetation might shift toward a more 3D (deep-rooted) water-use strategy despite the inherent NPP penalty, because the ecohydrological benefits (dry-season survival and enhanced rainfall recycling) become crucial. We will revisit this idea when examining past warm climate episodes.

*Scaling of NPP with Evapotranspiration: Within- vs. Across-Biome Patterns*
Evidence has been accumulating for decades that reveals a distinction between within biome and across biome behavior of the relationship between NPP and ET.

Analyses based on satellite pixel data (Zhang et al. 2016; Poveda, unpublished, 2025) reveal a nearly linear dependence of NPP on ET within biomes. This result is consistent with a near-constancy of Water Use Efficiency (WUE), the ratio of carbon fixed per unit water transpired (WUE = NPP/ET), within a given plant functional type. When structural traits such as leaf area, rooting depth, and canopy architecture are fixed, ET variations primarily reflect climatic forcing rather than ecosystem reorganization, yielding an approximately linear NPP–ET relation. Across biomes, however, structural and physiological traits vary widely, and WUE itself scales with ET through changes in vegetation density and its architecture, nutrient cycling, and soil development, consistent with a stronger dependence of NPP on ET. Such a stronger relationship has also been reported and confirmed.

Rosenzweig (1968) first reported a power-law dependence (NPP $\propto$ ET$^{1.69}$) across biomes, later confirmed by Milne (personal communication), who identified two spectral peaks near exponents 2 and 2.5. Budyko's (1973) continental-scale compilation, digitized in 2002 by Gupta and Poveda (explained in Hunt et al. 2024), similarly supports an exponent near 2, consistent with field data spanning nearly four orders of magnitude NPP and two orders of magnitude in ET, from ~0.5 g m$^{-2}$ yr$^{-1}$ at ET ≈ 20 mm yr$^{-1}$ to over 2000 g m$^{-2}$ yr$^{-1}$ at ET ≈ 1600 mm yr$^{-1}$. This cross-biome superlinearity reflects the compounding effects of increased water flux and ecosystem capacity: as ET rises, both plant activity and soil nutrient fluxes (linked to soil depth z $\propto$ (P – ET)$^{1.15}$) increase, amplifying productivity faster than linearly.

This nested scaling, linear within biomes but quadratic across them, fits naturally within the ecohydrological optimality framework. It reflects how local physiological constraints give way, at larger scales, to emergent ecohydrological organization that maximizes productivity across the coupled water–carbon continuum.

## 2.5 Evaluation of Evapotranspiration in Energy-Limited Environments by Physical Constraints

The optimal water balance we derived (Eq. 15a/15b) has a form that differs from conventional treatments of the water balance (Budyko's 1973 argument assumes ET/PET → 1 in very wet climates). However, two important points lend confidence to our formulation. First, at high aridity (PET/P $\gg$ 1), our model's prediction that Q = P – ET grows proportionally to P² (ET = P – 0.377 P²/PET makes Q $\propto$ P²) is supported by data. This has been verified through streamflow elasticity measurements: the median value of the logarithmic derivative of Q with respect to P is indeed 2 in dry regions, consistent with Q $\propto$ P². Second, the functional form of our ET solution in the humid limit can be derived from basic physical arguments. Kleidon et al. (2014) and Brutsaert (1982) independently showed that, in energy-limited conditions, one can expect an ET of the form ET = C · PET (with C < 1). They derived an expression for the constant C based on surface energy balance considerations:

$$C = \Delta / (\Delta + \gamma), \qquad (16)$$

where $\Delta$ is the slope of the saturation vapor pressure curve (Clausius–Clapeyron) at the surface temperature, and $\gamma$ is the psychrometric constant, which relates the increase in air vapor pressure to an increase in temperature when evaporation and sensible heat exchange are coupled; it is given by $\gamma = \frac{c_p P}{\lambda \varepsilon}$, with $c_p$ the specific heat of air at constant pressure, $P$ the air pressure, $\lambda$ the latent heat of vaporization, and $\varepsilon = 0.622$ the ratio of molecular weights of water vapor to dry air. Physically, C is less than 1 because some fraction of incoming solar energy goes into sensible heat flux into the ground rather than evaporating water. For typical Earth surface conditions, C is on the order of 0.6–0.75. The key point is that Brutsaert's physical treatment has the same functional form as our optimality, and that the value of C can be the same as our α, though it varies with temperature.

But because our ecological optimality factor α is derived from principle rather than explicitly from temperature, α itself is *constant* in our model (either ~0.623 or ~0.813). The physical C, however, increases with temperature (since $\Delta$ grows with T). Notably, at the present-day global mean temperature (~15 °C), the Kleidon/Brutsaert calculation gives C ≈ 0.62. This is essentially identical to our α = 0.623 for the 2D shallow-root optimum. Meanwhile, to get C ≈ 0.813 (the 3D deep-root optimum) from equation (16), one needs a much higher temperature – approximately 33 °C. Thus, the two critical α values from the ecohydrological model correspond to specific temperatures ~15 °C and ~33 °C at which the purely physical energy-balance constraint happens to allow the same fraction of ET *in humid climates*. This correspondence derived for energy-limited systems (PET/P < 1) extends to water-limited systems only insofar as any Budyko theory connects these two extremes, which may imply that the aridity index must tune separately for different temperatures, providing a possible basis for explaining complications in using simple Budyko equations for interpreting trajectories of catchments in Budyko space (Reaver et al. 2022).

This close correspondence between physical ET and optimality suggests that the prevalence of the 2D optimal water use in today's world may simply be a result of Earth's current temperature (~15 °C). In other words, contemporary ecosystems might appear to validate the optimality hypothesis (with α ≈ 0.62 everywhere) not because life always self-organizes to α = 0.623, but because we live in a climate where the physical limit coincides with that optimal value. However, this "coincidence" can be turned around and used as a clue to interpret the past global average temperature (GAT), as well as the present distribution of optimal temperatures for productivity around the globe**.** It also implies that, when Earth's climate was warmer (near 33 °C global mean), a different regime (analogous to α ≈ 0.813 optimality) could have been naturally favored. In the sections that follow, we will argue that throughout the Phanerozoic, the global climate tended to hover near these two preferred states (around 15 °C or 33 °C) and transitions between them were mediated by the biosphere and hydrologic feedbacks.

To summarize, the agreement between our α = 0.623 optimum and present-day data may be partly coincidental, reflecting Earth's current mean temperature. Yet this concordance also suggests that the 15 °C and 33 °C benchmarks are not arbitrary, but rather mark the intersection between ecological optimality and the physical limits imposed by the climate system. Below, we use these insights to help explain why Earth's Phanerozoic Global Average Temperature (GAT) has been largely bounded between 15 °C and 33 °C.

**2.6 Synthesis: Silicate Weathering, Plant Growth and Ecosystem Adaptation**

Overall, we emphasize several key points from the above theoretical considerations before comparing with empirical data:

1. **Optimal Greenhouse for Life:** While a certain level of greenhouse warming by $CO_2$ was essential to keep early Earth habitable (without any greenhouse effect, Earth would be a frozen ball), an optimal range of greenhouse conditions exists for the biosphere. Too much warming (e.g., near 33+ °C global mean) renders most of the planet inhospitable to complex life, whereas too little (below freezing) obviously does the same. This optimal climate for life has not been fixed; it likely shifted over geologic time and may not coincide with what is ideal for human civilization. The biosphere, through evolution and feedbacks, may have adjusted Earth's climate toward its own optimum, a perspective broadly consistent with Gaia hypotheses (e.g., Kleidon 2002).

2. **Weathering as a Thermostat and Its Limits:** Chemical weathering of silicate rocks is a major long-term sink of $CO_2$ and acts as a stabilizing feedback (the "weathering thermostat"). However, to model past climates accurately, one must capture the true nature of weathering kinetics. Field weathering rates are up to ~$10^6$ times slower than laboratory rates for the same minerals, due to transport limitations. We showed that using a transport-limited weathering theory (as opposed to an Arrhenius temperature-dependent rate law) removes many inconsistencies between models and data. For example, previous carbon-cycle models (e.g., Mills et al. 2019) had to introduce two *ad hoc* tweaks to reproduce past temperature estimates: they scaled reaction rates with runoff (acknowledging water control) and assumed that chemical reactions "saw" only tropical temperatures (which are more stable) rather than global averages. In effect, they were suppressing the unrealistic high temperature sensitivity in their Arrhenius formulation. By contrast, our approach naturally yields much lower sensitivity of weathering to global temperature, but high sensitivity to runoff and erosion, which aligns with both proxy evidence and geological intuition. As a result, we avoid the contradictions that plagued earlier models and gain more reliable insights into the carbon cycle's operation through time.

3. **Vegetation-Climate Feedbacks:** The terrestrial biosphere (vegetation) exerts powerful feedbacks on climate. Plants draw down $CO_2$ via photosynthesis, but at the same time they release water vapor via transpiration. Since water vapor is a potent (though short-lived) greenhouse gas, an exuberant biosphere can actually warm the climate even as it lowers $CO_2$, effectively trading one greenhouse gas for another. On short timescales, the warming from extra humidity can outweigh the cooling from $CO_2$ uptake (unless the carbon is locked away for very long periods in biomass or sediment). This means that an increase in plant productivity tends to humidify and warm the atmosphere in the near term. This tendency is accentuated by the effects of plants on the surface albedo (green Earth absorbs more sunlight). Effects of plants on clouds from increasing moisture may have either warming or cooling effects. Other aspects of vegetation feedback are cooling: for instance, more precipitation (driven by more vegetation) enhances silicate weathering and carbon sequestration in soils and sediments. The net effect of the biosphere on climate thus depends on a balance of positive and negative feedbacks – water-driven warming vs. carbon-driven cooling – and this balance can shift over time. However, a rise in Global Average Temperature (GAT) beyond 33 °C would demand adaptation to higher water-use efficiency and could render vast regions nearly devoid of vegetation. Such desertification would, in turn, amplify surface albedo and dry the atmosphere, producing a strong negative feedback on temperature, as seen today in the Sahara, CIMSS https://cimss.ssec.wisc.edu/wxwise/homerbe.html). Below, we consider how these feedbacks might have played out during episodes when Earth's climate pushed against the ~15 °C or ~33 °C bounds.

4. **Time Lags in Biospheric Responses**: Life's influence on climate is not instantaneous, especially when considering major evolutionary innovations. Plants spread over land partly by diffusion of seeds and spores, which is fast over short distances but very slow over continental length scales. Also, when plants evolve a new capability (e.g. producing lignin), it can take a long time for decomposer organisms (microbes, fungi) to evolve complementary pathways to break down that new material and close the carbon cycle loop. There are thus inherent delays in the Earth system between a biological innovation and the full climate/chemical equilibration. Hunt et al. (2025c) proposed that the timescale for an ecosystem to adapt and spread globally after a major innovation can be estimated using the same optimal transport scaling as for roots. Extrapolating from observed scales (up to ~100 kyr over ~10 km) to a whole continent (~5000 km) suggests on the order of ~80 Myr may be needed for an entirely new plant-decomposer network to proliferate worldwide. This hypothesis is speculative, but provides a quantitative framework for understanding why, for example, there were tens of millions of years of rising atmospheric $O_2$ and global cooling after the evolution of deep-rooted forests (before decomposers caught up), or why the first land plants triggered prolonged increase in atmospheric $O_2$ and a deep ice age (Mills et al. 2023). In the following sections, we will see that the timing of climate and atmospheric changes after key evolutionary events (Ordovician land plant invasion, Devonian rise of trees, late Miocene expansion of C4 plants) roughly matches these predicted adaptation timescales. Such lags imply that Earth's climate, carbon cycle, and biosphere were often in a state of transient imbalance, only slowly moving toward a new equilibrium after each major perturbation.

With these points in mind, we proceed to compare our theoretical expectations with the actual Phanerozoic data on global temperature, weathering, and biotic evolution.

## 3. Comparison with Global Data
### 3.1 Comparison with Silicate Weathering Rates: Modeled vs. Observed
Berner (1999) noted an intriguing pattern in the early Phanerozoic: atmospheric $CO_2$ levels and global temperatures appear positively correlated (high $CO_2$ accompanying warm climates), while at the same time atmospheric $O_2$ levels show an inverse trend: less $O_2$ in those warm, high- $CO_2$ times. This combination strongly implicates the biosphere, particularly land plants, in driving climate changes. High $CO_2$ and warmth with low $O_2$ suggests that plants were not very active (since vigorous plant growth would consume $CO_2$ and produce $O_2$, whereas later in the Phanerozoic, as plants proliferated, $CO_2$ tended to drop and $O_2$ rose, cooling the climate. Thus, the evolution of photosynthetic land ecosystems is tightly linked to global temperature regulation.

Silicate weathering, on its own, is often credited with preventing a runaway greenhouse (like on Venus) by removing $CO_2$ as temperatures rise. However, the classical formulation (Urey reaction with Arrhenius kinetics) misses the crucial role of water. Without liquid water, weathering grinds to a halt, no matter how high $CO_2$ or temperature get. This is evident in extreme scenarios: during "Snowball Earth" episodes in the Neoproterozoic, continental weathering essentially ceased because everything was frozen, allowing $CO_2$ to build up to extraordinary levels until deglaciation occurred. In more moderate situations, like arid climates, limited precipitation similarly throttles weathering rates. As discussed, after even days to months of exposure, fresh mineral surfaces become dependent on water flow to continue reacting. Transport limitations make weathering self-limiting in the absence of water flux, and prolonging the time or distance of transport further biases the limitation toward transport (rather than intrinsic kinetics).

Climate modelers have realised that representing silicate weathering as a mere thermostat in climate models oversimplifies reality, resulting in pronounced discrepancies with Earth's climatic evolution. Both Gibbs et al. (1999) and Mills et al. (2019) recognized that they needed to include runoff (Q) explicitly to get meaningful results. As Mills et al. (2019) state: "Global rates of silicate weathering are estimated in box models from the global average surface temperature (via Arrhenius) and scaling by the global rate of runoff." This was a step in the right direction, yet problems remained. Even with Arrhenius-type temperature dependence, the models produced temperature fields that were overly stable, showing far less variability than indicated by proxy reconstructions. The only way Mills et al. could make their model fit the Phanerozoic record was to assume that the effective temperature driving weathering was the relatively unchanging tropical temperature, rather than the more variable global mean. In essence, they had to remove the strong global T-dependence entirely to match observations (see figure 3). Gibbs et al. (1999) similarly found little evidence of a temperature-driven weathering feedback in their analysis. Instead, they highlighted the importance of paleogeography and climate: when continents were clustered into a supercontinent (e.g., Pangea, Permian period), conditions were drier and total $CO_2$ weathering fluxes were lower, whereas dispersed continents (e.g., during times of a vast Tethys Sea) were wetter with higher weathering fluxes. Silicate vs. carbonate weathering, in their results, both mirrored these aridity/humidity trends. These findings led Mills et al. (2019) to warn that any model which cannot capture the known climate swings (e.g., the extreme cooling of the Late Ordovician or the shifts in the Cenozoic) should not be trusted for deep-time predictions.

By applying Equations (3)–(7) from our theory, we solve both acknowledged problems in one framework: (i) the weak sensitivity of silicate weathering to temperature emerges naturally (since beyond short timescales, weathering is transport-limited, not kinetics-limited), and (ii) the strong sensitivity to runoff (and by extension, to factors like continental humidity and relief) is explicitly included. We also add the dependence on relief (mountain uplift, etc.) that others have noted qualitatively but not captured in simple models. Thus, our approach can reconcile model predictions with the data without the need for *ad hoc* adjustments.

The significance of water availability (or lack thereof) for catastrophic climate events is exemplified by the end-Permian extinction. In an interview (Editors' Vox, Eos) discussing (Hunt and Sahimi, 2017), we responded to a question about combining theory and application by saying: "Very simply put: no water, no weathering, no $CO_2$ drawdown, no cooling – thus extreme extinction event." During the late Permian, the supercontinent Pangea had vast arid interiors. Even though global warming typically increases overall precipitation, the distribution was such that much of Pangea remained extremely dry. The normal negative feedback (higher T → more rain → more weathering → cooling) was largely short-circuited because there was so little rain where it was needed. As a result, $CO_2$ stayed high and temperatures soared to nearly 33 °C (in fact, in some reconstructions they just reach 33 °C around the Permian–Triassic boundary). This unmitigated warmth contributed to what is recorded as the most severe mass extinction in Earth's history. Our perspective underscores that it was not just the massive $CO_2$ emissions from Siberian Traps volcanism that made the Permian extinction so severe, but also the climatic context of extreme continental aridity which prevented the Earth system from buffering that $CO_2$ spike via weathering.

After the extinction, the Earth did eventually recover, but the biosphere's recovery in the Early Triassic was inordinately slow. Estimates suggest it took on the order of 5-10 Myr for ecosystems to re-establish fully (Gradstein et al. 2012; Tong et al. 2007). One likely factor was the continued aridity in the early Triassic (Cui & Cao, 2023), which would have kept weathering rates low and $CO_2$ levels high even after volcanic emissions waned. Other research has debated whether the terrestrial extinction was truly a wipe-out or more of a turnover (e.g., Romano et al. 2013; Xiong & Wang 2011; Nowak et al. 2019). If many plant lineages survived but were replaced gradually ("adaptive replacement"), that implies the land biosphere was rebuilding from sparse remnants under harsh climates. In any case, with $CO_2$ remaining high and very slow weathering drawdown (due to lack of moisture), it fell partly to the recovering biosphere itself to eventually draw down $CO_2$ (through organic carbon burial and modest weathering once climates dampened a bit). Only when the intense volcanic carbon input ceased and enough time passed for even slow weathering to integrate to large effects did the climate fully stabilize out of this hot state.

As mentioned earlier, an extreme case of "no water, no weathering" occurred during Neoproterozoic Snowball Earth events. Continental surfaces were largely ice-covered, effectively eliminating silicate weathering for millions of years. This allowed $CO_2$ from volcanoes to accumulate to perhaps >0.1 bar (~100,000 ppm) – levels high enough to eventually melt the global ice once the radiative forcing crossed a threshold (Hoffman et al., 1998). Our framework provides a straightforward explanation: with weathering almost wholly suppressed, $CO_2$ can rise unabated until the climate flips state. This underscores again that, without water, the thermostat is broken.

Having established how incorporating hydrologic and biotic controls improves our understanding of weathering and climate, we next examine whether the actual global temperature record of the Phanerozoic exhibits the patterns predicted by our ecohydrological optimality model. In the comparisons below, we consider three studies examining continental- to global-scale variations in vegetation optimal temperatures under present conditions, together with six published reconstructions of Phanerozoic global temperature (Mills et al., 2019; Scotese et al., 2021; Shaviv et al., 2023; Judd et al., 2024), and one additional study focused specifically on the most recent part of the Phanerozoic, the Cenozoic.

### 3.2 Comparison with Current Optimality Temperatures

If the 15 °C and 33 °C benchmarks are to be relevant to Earth's history, we should be able to see some evidence for their relevance at the present time, as well. Indeed, in Australia, "Ecosystem [for maximum productivity] $T_{opt}$ ranged from 15 °C (temperate forest) to 32°C (tropical savanna-wet and dry seasons)" (Bennett et al., 2021). Of course, Australia is a warm and dry continent, not allowing for any significant test of conditions below 15 °C. More comprehensive datasets reveal very similar results, however. In Figure 1, we compare these temperature bounds with data at the global scale for the latitudinal dependence of the optimal temperature for vegetation gross primary productivity (GPP), as reported by Huang et al. (2019).

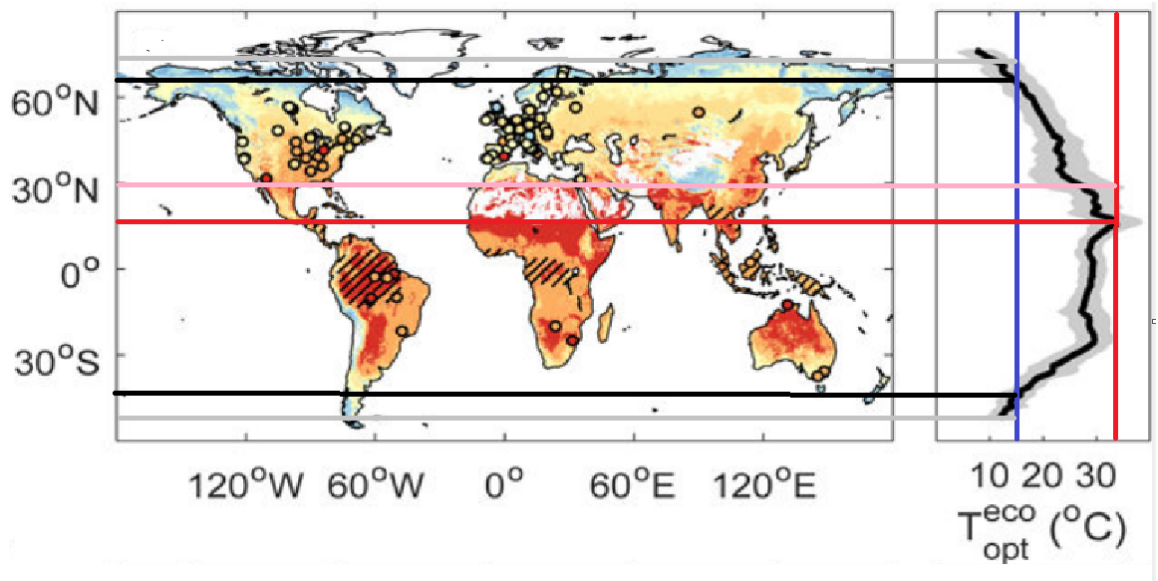

Figure 1. (right panel): average temperature for ecological optimality as defined by maximum gross primary productivity (GPP) as a function of latitude (solid black curve), as well as the spread in temperature values (grey) as defined by one standard deviation. (left panel): Their boundary between light yellow and light blue

represents 15 °C and their boundary between red and orange represents 30 °C. Our predictions of 33 °C and 15 °C are given by the vertical red and blue lines in the right panel. Note that the peak mean ecological optimal T at 33 °C, denoted by the red horizontal line, maps out the boundary between the Sahel and the Sahara. The pink horizontal line denotes the only other latitude, where one standard deviation above the mean ecological optimum T values exceeds 33 °C, and coincides with the northern boundary of the Sahara. The horizontal black lines delineate the range of latitudes where the average optimal temperature lies between the two proposed values 33 °C and 15 °C, while the horizontal grey lines delineate the latitudes where at least some individual values fall within that same temperature interval. The landmass that is without continental glaciation lies almost exclusively between the bounds at 15 °C. After Huang et al. (2019).

Figure 1 demonstrates the relevance of the ecological optimality temperatures 15 °C and 33 °C, corresponding to the 2D and 3D optimality predictions for the water balance to the range of latitudes without continental glaciation. As might be expected (Bennett et al. 2019), the mean ecological optimum 15 °C latitude of the southern hemisphere corresponds to the southern limit of Tasmania, while one standard deviation below the mean passes approximately through Cape Horn. Even in the tropics, the mean optimal temperature for productivity only reaches 33 °C, but does not pass it. Both Fang et al. (2024) and Bennett et al. (2021) suggest that ecologically optimal temperatures of all ecosystems show adaptability to rising temperatures, and predict continued increases in temperatures associated with optimal ecosystem gross primary productivity (GPP); however they may not be able to predict accurately whether the GPP associated with the optimal temperature actually increases across the 33 °C boundary. In our treatment, the demands on the root system would preclude such a continued increase. Note that the range of latitudes in which the observed upper envelope of the ecological optimum temperature reaches or exceeds 33 °C bounds the Sahara Desert, which exhibits net radiative cooling (CIMSS https://cimss.ssec.wisc.edu/wxwise/homerbe.html), in contrast to the tropics generally, which have a net positive radiation budget. As can be seen, Figure 1 is consistent with our general interpretation of feedbacks discussed in section 2.

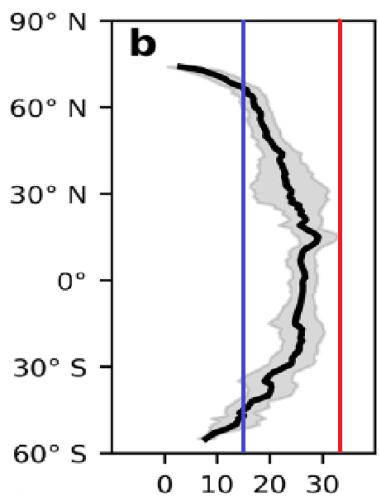

Figure 2. Ecological optimality temperature associated with maximum productivity as a function of latitude. Red line shows temperature associated with 3D optimality calculation of evapotranspiration (ET) blue line shows temperature associated with 2D optimality. In contrast to Figure 1 (Huang et al. 2019), where the ecological optimal temperature 33 °C corresponds to the mean value at 15 °N, the data of Pan et al. (2022) show that 33 °C lies approximately one standard deviation above the mean optimal temperature at the same latitude of 15 °N. After Pan et al. 2022.

Interestingly, in Figure 2, there appears to be a sudden increase in the slope of T vs. latitude at T = 15 °C in both hemispheres even though the latitudes at which the value of 15 °C shows up is near 50° in the southern hemisphere and near 65° in the northern. This slope increase may correspond to the onset of a positive feedback during cooling past GAT = 15 °C in some reconstructions of global climate.

## 3.3 Fluctuations in Phanerozoic Global Surface Temperatures
### 3.3.1 Empirical bounds in Phanerozoic temperature reconstructions

When we superimpose the two characteristic temperatures from our theory, 15 °C and 33 °C, onto the Phanerozoic temperature reconstructions (Figures 3–6), the relevance of these optimum bounds becomes immediately apparent. Across all datasets, Earth's inferred Phanerozoic global mean temperature rarely exceeds ~33 °C on the high side, and glacial episodes correspond to (mostly) short-lived drops below ~15 °C on the low side. The consistency of these bounds, despite differing methodologies and proxies, is striking. The repeated emergence of 33 °C as the maximum optimality temperature in the subtropics, and of 15 °C as the present-day global average near the edge of an ice age, is striking. This alignment, with 15 °C also marking the lowest ecological optimality temperatures near glaciated margins, is unlikely to be a coincidence.

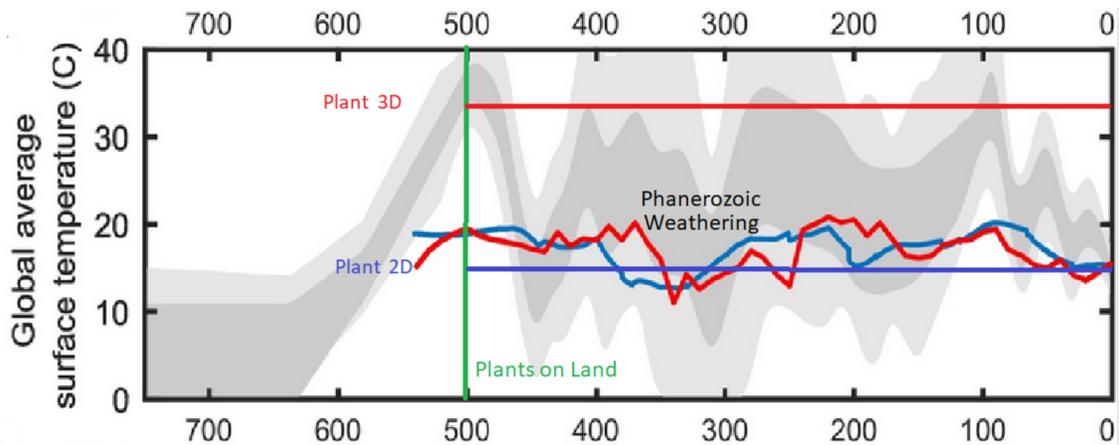

Figure 3. On the possible relevance of ecohydrological optimality to the global temperature record of the Phanerozoic (Mills et al. 2019). The dark grey corresponds to one sigma range of possible temperatures that is compatible with both of their two primary paleo-proxies. The horizontal blue and red lines correspond to the temperatures 15 °C and 33 °C from ecohydrological optimality. The two "Phanerozoic Weathering" curves (blue and red) correspond to model outputs from Mills et al. (2019) using the same silicate weathering framework, with the blue curve including a global runoff correction and tropical temperature forcing. Both exhibit similar variability, underscoring that even modified Arrhenius-type formulations remain too stable compared to proxy-inferred temperature fluctuations.

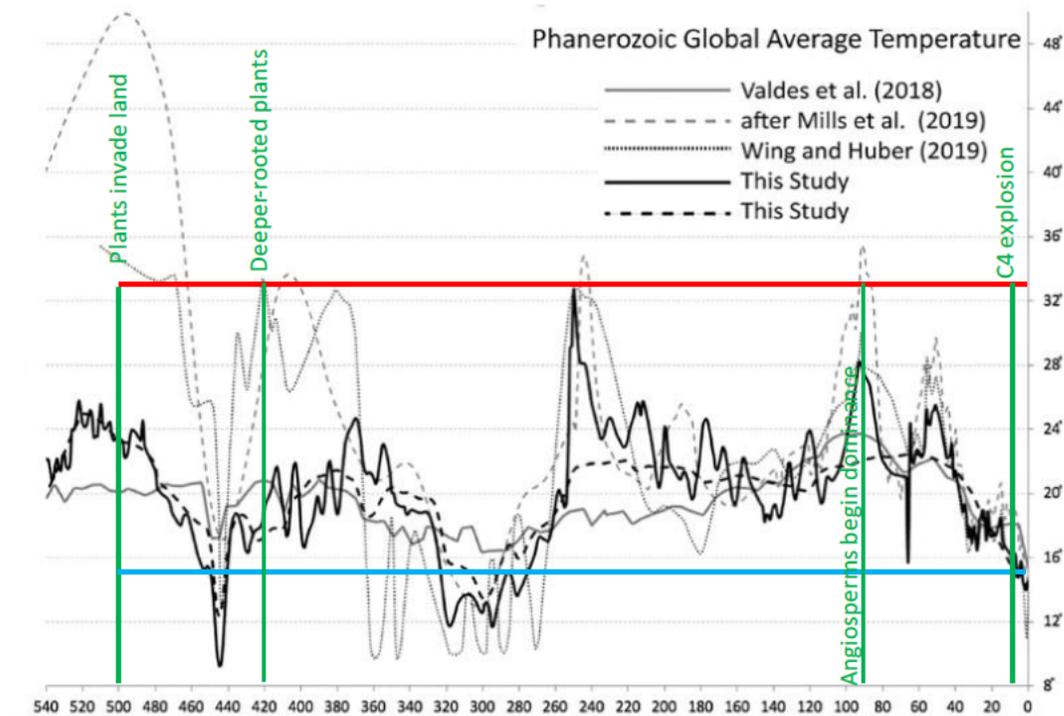

Figure 4. Comparison of T = 33°C and T = 15°C with the Phanerozoic temperature record of Scotese et al. (2021), who also show the reconstructions of Valdes et al. (2018), Mills et al. (2019) and Wing and Huber (2019). The two reconstructions of Scotese et al. (2021) are the solid black line and the thick dashed black line. The results are based on the simultaneous consideration of multiple proxies. Of the five reconstructions and three main periods of warming, only one data set exceeds 33°C by as much as 1 °C, once at 34 °C and 35 °C. But glacial episodes can penetrate the 15 °C boundary. Approximate times of key plant innovations are indicated.

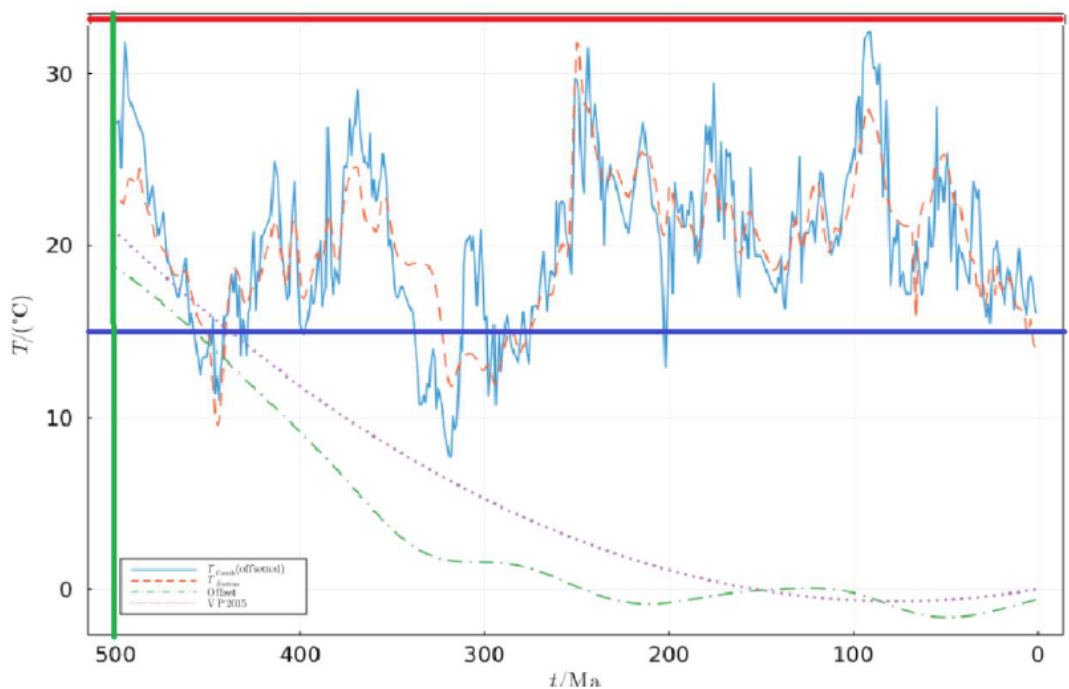

Figure 5. Shaviv et al.'s global T reconstruction (2023) in blue curve. Dashed red curve is from Scotese et al. (2021). The upper boundary of 33C is, as in the Mills et al. (2019) reconstruction, better represented than in Scotese et al. (2021). However, Paleozoic glacial episodes penetrate the lower boundary at 15C more often and more deeply.

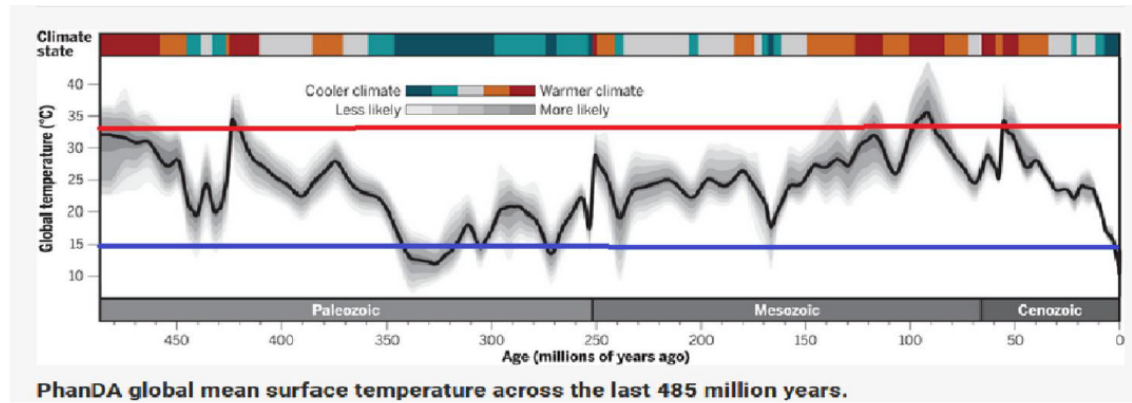

Figure 6. Phanerozoic temperatures as deduced by Judd et al. (2024). Here, both boundaries, at 15C and at 33C are plainly visible in the data, which include fewer of the higher frequency fluctuations.

Figure 3 (after Mills et al. 2019) shows a multi-proxy temperature range for the Phanerozoic (dark grey band) alongside the temperatures predicted by an unmodified silicate weathering model. The silicate weathering model fails to capture the real variability, but note how the empirical temperature range never goes far beyond ~15 °C to ~33 °C. Figure 4 (from Scotese et al. 2021) presents two reconstructions (solid and dashed black lines) based on large proxy compilations. Superimposed are other models, including (Mills et al., 2019), and horizontal lines at 15 °C and 33 °C. The 33 °C line effectively bounds the peak warm events, even the highest temperature spikes in any reconstruction linger just above or below 33 °C. On the cold end, glacial intervals (such as the end-Ordovician, late Devonian, late Carboniferous-Permian, and Pleistocene) clearly dip into the sub-15 °C range, showing that 15 °C often marks a threshold between nonglacial and glacial climates. Figure 5 (blue curve from Shaviv et al. 2023, red dashed from Scotese 2021) again highlights that, while various reconstructions differ in detail, none sustain temperatures above ~33 °C, and many Paleozoic glacial episodes are times when temperatures breached ~15 °C on the downside. Figure 6 (Judd et al., 2024) presents smoother, lower-frequency temperature variations with both a ceiling near 33 °C and a floor near 15 °C over 500 Myr. These visual comparisons suggest that our identified optimal points were indeed rarely exceeded.

**3.3.2 What about specific events? The Late Ordovician (Hirnantian) glaciation**
Consider the Late Ordovician Hirnantian glaciation (~450–440 Ma), one of the most extreme ice ages of the Phanerozoic. It occurred during a period that was otherwise expected to be warm and it coincided with a "tectonic cycle" phase (Wilson cycle) associated with dispersed continents and greenhouse conditions. Similarly, a lesser glaciation in the Late Devonian (around 372–359 Ma) happened when, by tectonic configuration alone, climate might have been warmer. These apparent anomalies prompted researchers to look for biological causes. Lenton et al. (2012) argued that the Hirnantian glaciation was triggered by the advent and spread of the first land plants, which would have drastically enhanced weathering and drawn down $CO_2$ (and also possibly released cloud-seeding compounds). The timeline fits: simple non-vascular plants began colonizing land around 500Ma (Morris et al. 2018; Puttick et al. 2018) and 480Ma (Mills et al. 2023), the latter of which would imply that within ~30 Myr the world plunged into an ice age. Mills et al. (2019) similarly observed that the two glaciations their model could not account for—the Late Ordovician and the current Cenozoic—corresponded to pivotal evolutionary events: the first terrestrial colonization and, much later, the emergence of $C_4$ vegetation. This reinforces the hypothesis that the biosphere was the wildcard tipping the climate scales in those cases.

### 3.3.3 Ordovician and Devonian intervals in context

We now focus on the Ordovician and Devonian intervals, interpreting them within the context of our framework. Following the initial colonization of land by plants (bryophyte-like flora) around 500 Ma (Morris et al. 2018; Puttick et al. 2018), global temperatures were very high: reconstructions show values around 33 °C, or even 36 °C in one model (Mills et al.) and ~24 °C in another (Scotese et al.) at ~480 Ma. By that time (~480 Ma) roughly 20 Myr of shallow-rooted plant presence had passed, and then a marked cooling commenced. From ~480 Ma to ~460 Ma, the average temperature fell from the low 30s to ~15 °C. In one reconstruction (Scotese's Fig. 3), the cooling seems to accelerate once ~15 °C is reached, and indeed by ~460 Ma the Hirnantian glaciation began. At the glacial maximum (~450–440 Ma), global mean temperatures may have bottomed out near 9 °C (in Scotese's curve) or ~12 °C (in Mills et al.'s). This was a short-lived but intense ice age. Interestingly, atmospheric $O_2$ levels were initially low (~3% in the Late Cambrian) and actually decreased slightly as plants first spread (500–480 Ma), possibly because early soils and ecosystems were sinks for $O_2$. But around the start of the cooling (480 Ma), $O_2$ began a steady rise, reaching ~20% by 420 Ma (Mills et al. 2023). In other words, after a ~20 Myr lag, the greening of land led to a massive oxygenation of the atmosphere (which itself has climate effects). Temperature recovered from the Hirnantian low by ~440 Ma, bouncing back above 15 °C, but $O_2$ kept climbing until ~420 Ma. Measured from the onset of significant plant invasion (~480 Ma) to the end of its climate impact, we see about a 40 Myr span of cooling (to 440 Ma) and a ~60 Myr span of rising $O_2$ (to 420 Ma). Either way, on the order of 50 Myr of profound climate and atmospheric change can be attributed to the introduction of land plants. This aligns with our earlier argument that ~80 Myr might be needed for the Earth system to adapt to such a fundamental change (here the delay was ~20 Myr to start cooling, then ~40 Myr to fully recover temperature, and a total of ~60 Myr for $O_2$ equilibration).

### 3.3.4 Devonian-Carboniferous transition and oxygen rise

Let us now consider the Devonian-Carboniferous period, when deep-rooted woody plants (trees) emerged. The first forests of taller trees and true roots appeared by ~420 Ma (Silurian-Devonian boundary). Our theory predicts another disruption as expanding forests accelerated weathering and carbon sequestration, owing to the buildup of lignin-bearing wood that was initially resistant to decay. Indeed, another long cooling trend develops from roughly 375 Ma to 275 Ma. Figure 4 (Mills) shows ~33 °C at 400 Ma falling to ~15 °C by 300 Ma, and Figures 5–6 (Scotese et al. 2021, Shaviv et al. 2023) show ~29 °C at 375 Ma falling to ~12 °C by 275 Ma. There was a glaciation in the Late Devonian (Famennian, ~372–359 Ma) reported by geological data (Streel et al., 2000), though interestingly it does not manifest clearly in the temperature curves as it may have been regionally constrained or masked by data gaps. During 375-275 Ma, atmospheric $O_2$ skyrocketed from ~15% to over 30%, the highest oxygen levels of the Phanerozoic (Mills et al. 2023). This is the time of the Carboniferous "coal age," when immense volumes of organic carbon were buried as coal. Tree height went from a few meters (Dilcher et al. 2004) in the early Devonian to 30–50 m by the late Carboniferous (Arborescent lycopsids, ferns, seed ferns, and the first conifers), indicating the development of dense, extensive forests. Lignin, a complex woody tissue component, appeared and was something new for decomposers to tackle. As Berner (1999) noted, the Carboniferous coal deposition rate was unparalleled, suggesting that organic carbon (and oxygen byproduct) accumulated because decomposition lagged behind production. Fungi capable of breaking down lignin (white rot fungi) did not evolve until the Late Carboniferous or Permian, according to some studies, which supports this idea. The net effect was that, for on the order of 100 Myr, oxygen production outpaced consumption, driving $O_2$ to perhaps ~35% (Mills et al. 2023).

Eventually, probably by the Permian, decomposer food webs evolved to catch up, breaking down lignin-rich material more efficiently and contributing to the fall in atmospheric $O_2$ after its peak.

Let us compare the Devonian-Carboniferous event to the Ordovician one. The lag between innovation (woody plants ~420 Ma) and climate effect (cooling evident by ~375 Ma) is about 45 Myr, longer than the ~20 Myr lag for the first plants. The duration of elevated $O_2$ (~100 Myr) is also longer than the ~60 Myr in the earlier case. The interval between the initial biological innovation and the termination of the associated cooling phase, approximately 420 Ma to 300 Ma (~120 Myr), is relatively long. However, if the onset of the first Devonian glaciation is taken to occur roughly 60 Myr after the appearance of vascular trees, this lag is comparable to that observed in the Ordovician case (500 Ma to 440 Ma, ~60 Myr between the emergence of land plants and the first major glaciation). In both cases, the first significant glacial episode follows the ecological innovation after a delay on the order of several tens of millions of years. Although the precise timing remains uncertain, this correspondence is consistent with the ~80 Myr adaptation timescale predicted by Hunt and Manzoni (2015) for the coevolution of coupled plant-decomposer networks. These temporal estimates, on the order of 60 to 100 Myr, are consistent with the expected timescales discussed above and represent a reasonable match given the inherent geological uncertainties. The extended duration of the Devonian-Carboniferous high-$O_2$ interval may reflect the exceptional biochemical challenge posed by lignin degradation, which could have prolonged the adaptation period of decomposer networks. In either case, the correspondence is striking: the two largest declines in global average temperature and the most pronounced rises in atmospheric oxygen both coincide with the two major evolutionary transitions in terrestrial vegetation.

After ~275 Ma (mid-Permian), the trend reverses. Global temperature rose rapidly from the late Permian into the Early Triassic, while atmospheric $O_2$ dropped from ~30% toward ~15–20%. This is consistent with the idea that, eventually, decomposers (microbes, fungi) evolved to break down much of the accumulating plant biomass (especially lignin-rich wood). Massive peat deposits that had formed (future coal) were increasingly oxidized or their formation slowed. The carbon that had been locked in vast coal swamps started to find its way back into the atmosphere as $CO_2$, warming the climate. And then came the Siberian Traps eruptions (~252 Ma), which dumped enormous $CO_2$ into the system on top of that. The climate at the end-Permian nearly reached the upper constraint (~33 °C) but notably, based on reconstructions (and as we pointed out from Shaviv et al. 2023 and Judd et al. 2024, data), appears not to have exceeded it. The biosphere was severely disrupted as many terrestrial plant lineages went extinct or were replaced. But it was not entirely eliminated, as some resilient vegetation persisted into the Triassic. Retallack (1995) described this interval as a "Dead Zone" followed by the emergence of a weedy, opportunistic flora. This residual vegetation may have been the critical factor preventing a complete runaway greenhouse. As discussed earlier, approaching a global average temperature of 33 °C likely activates strong negative feedbacks that stabilize further warming. These negative feedbacks include widespread plant dieback, increased surface albedo, and atmospheric drying. The fact that global temperatures did not exceed 33 °C even during this extreme event is consistent with that self-limiting mechanism.

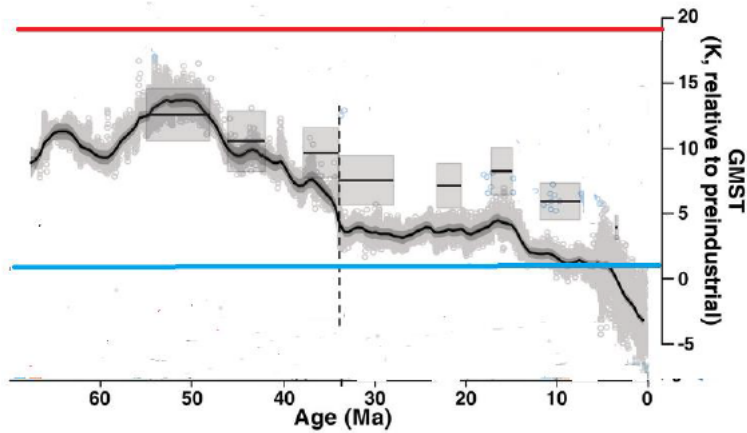

Figure 7. Comparison of the values of T = 15°C (1 degree above pre-industrial 14°C) and T = 33°C (19°C above pre-industrial) with Cenozoic temperatures (Hönisch et al. 2023). Cenozoic $CO_2$ Proxy Integration Project (CenCO$_2$PIP) Consortium. In the Scotese et al. (2021) reconstruction, the Hirnantian glaciation around 460 Ma shows a sharp cooling once global temperatures fall below ~15 °C; a comparable rapid decline occurs again near 4 Ma, suggesting that crossing this threshold may systematically trigger accelerated cooling.
s

### 3.3.5 Threshold behavior near 15 °C in the late Cenozoic

Finally, we turn to the late Cenozoic, focusing on the Neogene and, in particular, the Pliocene-Pleistocene transition. From the mid-Miocene (~17 Ma) through the Pliocene (~5 Ma), Earth's global average temperature exhibited a sustained cooling trend, reaching approximately 15 °C by around 3-4 Ma. It was at this threshold that the first large Northern Hemisphere ice sheets formed, marking the onset of the Mid-Pliocene to Pleistocene glaciations. High-resolution Cenozoic records (Figure 7) show that, once global temperature fell below ~15 °C, the rate of cooling accelerated, culminating in the onset of the Quaternary ice-age cycles. This mirrors what happened at ~460 Ma and ~360 Ma in the Paleozoic. If ice-albedo feedback were the only cause of glaciation, differences in continental positions should have led to different onset temperatures. The Hirnantian glaciation around 440 Ma was centered in Gondwana, while the Pliocene-Pleistocene glaciation around 3 Ma began in the Arctic and Greenland. Yet both appear to start near the same ~15 °C threshold. This recurrence points to a common mechanism, likely involving the biosphere. Even today, a mean growing season ecological (productivity) optimal temperature of about 15 °C coincides with the periglacial boundary in the Arctic, separating vegetated from glaciated regions.

### 3.3.6 C$_4$ Expansion as a Late Cenozoic Cooling Feedback

Why does cooling accelerate below 15 °C? Our feedback analysis in sections 2.5 and 2.6 provides a possible answer: below 15 °C, plants struggle to maintain the optimal water cycling, leading to reduced evapotranspiration (ET) and higher albedo, thus a positive feedback to cooling.

In the Pleistocene's case, another factor came into play: the evolution and expansion of C4 grasses. Between ~8 Ma and 5 Ma, C4 photosynthesis (which had existed in some form earlier) became ecologically dominant in grasslands around the world (the "C4 revolution"). C4 plants are an evolutionary response to low $CO_2$ and aridity; they photosynthesize more efficiently under those conditions, losing less water per unit carbon fixed (higher water-use efficiency). By ~5 Ma, C4 grasses spread widely, transforming savannas and grassland ecosystems (Osborne & Beerling, 2007). Though C4 species are only ~3% of plant species, they account for about 20–30% of global net primary productivity (NPP) (mostly tropical grasses, including crops like maize and sugarcane). The effect of this transition on climate would be subtle but directional: C4 plants, by conserving water, draw down relatively more $CO_2$ from the atmosphere for the amount of water vapor transported. This tilts the balance toward cooling because, previously, C3 plants would release more $H_2O$ for a given $CO_2$ uptake. In a drying, cooling world, C4 plants essentially reinforce the cooling and drying, a positive feedback on cooling. We propose that the late Miocene expansion of C4 vegetation provided an extra push in the direction of the modern icehouse, beyond the tectonic and oceanic changes that were occurring. The atmospheric $O_2$ record does not show a huge change in the last 20 Myr (perhaps a small peak ~10 Ma per Mills et al. 2023), so the fundamental capacity of plants to affect $CO_2$ / $O_2$ did not change drastically. What changed was how plants used water versus carbon, and that can alter the $H_2O$ / $CO_2$ greenhouse balance. In essence, the biosphere found a new way (C4 photosynthesis) to cool and dry the planet without reducing its overall productivity.

One could argue that, absent human interference, this cooling might have continued. We are currently in a brief interglacial (Holocene) punctuating an ice age, and our fossil fuel emissions have raised $CO_2$ abruptly. But in a hypothetical future with no humans, perhaps $CO_2$ would have slowly kept declining (via weathering and organic carbon burial) and C4 grasslands might expand further, making the world even cooler. Unlike the Carboniferous lignin scenario, where eventually microbes evolved to consume the new carbon reservoir (breaking the negative feedback), there may be no analogous "decomposer" for more efficient photosynthesis. In other words, it is not obvious what natural adaptation would counteract the advantage of C4; it might be a permanent shift helping life keep $CO_2$ low as the sun's output gradually increases (a very slow, long-term warming influence). This is speculative, but it underscores the idea that the biosphere's innovations can have one-way effects on climate that persist until another innovation changes the game.

**4. Conclusions**
We have presented our central hypothesis that the long-term evolution of Earth's climate is bounded and modulated by ecohydrological optimality, that is, by the constraints imposed by water availability on plant function and by the optimal water-carbon balance in terrestrial ecosystems. We have proposed that these biological limits correspond to two characteristic global temperatures: roughly 15 °C, below which the biosphere struggles to sustain evapotranspiration and global water cycling, and 33 °C, above which vegetation suffers heat and water stress. These thresholds define a climatic corridor within which life maintains its productivity and, through feedbacks, stabilizes the planet's surface temperature.

We have explained that the physical mechanisms underlying this framework are rooted in the coupled behavior of the hydrological cycle and the biosphere. At the warm end, high temperatures and water stress reduce vegetation cover, increase albedo, and dry the atmosphere, producing a negative feedback on warming. At the cold end, declining evapotranspiration and enhanced reflectivity reinforce glaciation and accelerate cooling. Over geologic time, these hydrological feedbacks interact with the carbon cycle, particularly through their control on silicate weathering and organic carbon burial, to produce the quasi-stable temperature range observed through the Phanerozoic.

To test this hypothesis, we examined ten studies published between 2019 and 2024, including three on modern conditions (Bennett et al., 2021; Huang et al., 2019; Pan et al., 2024), six Phanerozoic-scale reconstructions (Valdes et al., 2018; Wing and Huber, 2019; Mills et al., 2019; Scotese et al., 2021; Shaviv et al., 2023; Judd et al., 2024), and one focused on the Cenozoic (Hönisch et al., 2023). All ten datasets independently confirm the recurring significance of the 15 °C and 33 °C thresholds, whether in the modern distribution of growing-season temperatures or in the deep-time temperature record. These findings support our view that life, water, and climate form an integrated feedback system that regulates Earth's long-term thermal state. Major evolutionary transitions, such as the colonization of land by plants, the rise of deep-rooted forests, and the emergence of $C_4$ photosynthesis, represent distinct reorganizations of this system, each introducing new feedbacks that reshaped global temperature and atmospheric composition.

Our framework predicts that future biospheric innovationsor human interventions, that alter evapotranspiration, albedo, or the global water-carbon balance, could again shift Earth's climate between quasi-stable states. On geological timescales, such shifts have been driven by evolutionary innovations like the rise of vascular plants, forests, and $C_4$ photosynthesis, each reorganizing the coupling between water, carbon, and energy fluxes. On shorter, anthropogenic timescales, similar principles apply: large-scale deforestation, irrigation, land-cover change, and atmospheric $CO_2$ enrichment already modify the partitioning of latent and sensible heat fluxes, influencing regional precipitation and cloud feedbacks. These changes can either amplify or counteract global warming, depending on how they alter evapotranspiration efficiency and surface reflectivity.

Refining this hypothesis therefore requires models that explicitly represent biophysical feedbacks between vegetation dynamics, surface energy balance, and the carbon cycle at both millennial and centennial scales. Improved formulations of silicate weathering, the inclusion of episodic forcings such as Large Igneous Provinces or bolide impacts, and tighter reconciliation among paleotemperature proxies will strengthen long-term predictive skill. Testing these mechanisms at shorter timescales using satellite data, ecohydrological measurements, and integrated Earth system simulations can show how rapidly biospheric processes adjust to changing conditions.

In this broader view, the biosphere emerges not as a passive boundary condition but as an active regulator of climate stability. Understanding how vegetation controls the partitioning of water and energy today will illuminate how the same principles have governed Earth's resilience across the Phanerozoic—and how they may determine the planet's trajectory in the Anthropocene and beyond.

**Acknowledgments**
The authors wish to acknowledge the fundamental contribution of Gabriel Katul (Hunt et al. 2025c) linking the physical theory of Brutsaert (1982) with ecohydrological optimality (Hunt et al. 2021), thus permitting the extension to ecohydrological influence on Earth's temperature. D.S. was partially supported by the National Natural Science Foundation of China (Grant No. T2350710802 and No. U2039202), Shenzhen Science and Technology Innovation Commission Project (Grants No. GJHZ20210705141805017 and No. K23405006), and the Center for Computational Science and Engineering at Southern University of Science and Technology.